\documentclass[epj,final]{svjour}
\usepackage[dvipdfm]{graphicx}
\begin{document}
\title{A Simple Model for the Checkerboard Pattern of 
Modulated Hole Densities in Underdoped Cuprates}
\titlerunning{A Simple Model for the Checkerboard Pattern
in Underdoped Cuprates}

\author{T.~M.~Rice\inst{1,2}%
  and Hirokazu Tsunetsugu\inst{2}%
}                     
\authorrunning{T.~M.~Rice and H.~Tsunetsugu}

%
\institute{%
Institut f\"ur Theoretische Physik, ETH-H\"onggerberg,
CH-8093 Z\"urich, Switzerland \and
Yukawa Institute for Theoretical Physics, Kyoto University,
Kyoto 606-8502, Japan}

\date{Received: \today}
%
\abstract{
A simple model is proposed as a possible explanation for 
the checkerboard pattern of modulations in the hole 
density observed in recent tunneling experiments on 
underdoped cuprates.  Two assumptions are made; 
first, an enhanced hole density near the acceptor 
dopants and secondly short range correlations in 
the positions of these dopants caused by their electrostatic 
and anisotropic elastic interactions.  
Together these 
can lead to a structure factor in 
qualitative agreement with experiment.  
} 
\maketitle

Recent studies which measured the tunneling conductance into
underdoped cuprates found an unexpected spatial modulation 
at higher bias voltages with a 
4$\times$4 periodicity.  The most comprehensive studies \cite{hanaguri} 
have been made on the
Ca$_{2-x}$Na$_{x}$CuO$_{2}$Cl$_{2}$ compounds where a good surface can be
prepared deep into the underdoped region.  These scanning tunneling
microscopy (STM) studies have led to interpretations of the data as
signals of superlattices of single holes or hole pairs 
\cite{anderson,chen,fu,vojta,tesanovic}.  Since these
materials are Mott insulators doped randomly with acceptors it is not
clear if superlattice ordering can persist in the presence of 
the random electric fields of the acceptor ions.  
In this letter we will explore
an alternative explanation that starts from the assumption of 
enhanced hole densities
localized in the vicinity of acceptors and examines the consequences
of short range correlations in the positions of the acceptors, e.g. 
the Na-ions in Ca$_{2-x}$Na$_{x}$CuO$_{2}$Cl$_{2}$.   

Our first assumption of enhanced hole densities close to 
the acceptors is consistent with the standard behavior of 
doped semiconductors.  These undergo a Mott transition when 
the dopant concentration is reduced and hole wavefunctions 
become localized around acceptors.  
Here we make a simple assumption of an enhanced hole density 
and explore the consequences.
Then we simulate the positions of
the Na ions under the combined influence of inter-ion Coulomb
repulsion and elastic forces, using a Monte Carlo algorithm.
Together these two effects determine the  form of
the modulations in the hole density and so the STM structure factor.
As we shall see the calculated structure factor shows qualitative
agreement with the STM experiments.

\begin{figure}[b]
\begin{center}
\end{center}
\caption{Projected view of crystal structure of
Ca$_{2-x}$Na$_{x}$CuO$_{2}$Cl$_{2}$ around a
doped Na$^{+}$ ion.  
Black and gray balls represent Cu and Ca ions.
An electron (or hole) injected to Cl 2$p_z$-orbital 
above site 0 is transferred to one of the Cu sites 1-4.  
} 
\label{fig:atoms} 
\end{figure}

The outermost layer of the crystal consists of interpenetrating
square lattices of Cl and (Ca$_{1-x/2}$Na$_{x/2}$) ions. 
The STM tip will couple most effectively to the outermost
wavefunctions of the surface layer which should be the 3$p_z$ orbitals
of the Cl ions.  The final state of the injected electron (+ve bias)
is in the hybridized antibonding band of 3$d_{x^2-y^2}$ orbitals
centered on Cu sites and the 2$p_{x,y}$ O-orbitals.   Since the Cl ions
lie directly above the Cu-sites there is no matrix element to tunnel
into the state with $d_{x^2-y^2}$-symmetry centered of the Cu-site
directly below.   There will be however a finite matrix element to
couple into the 4 nearest neighbor Cu sites generated 
through the overlap of the
3$p_{z}$-Cl and  2$p_{x,y}$-O states on neighboring Cl-O pairs.  This
situation is similar to the case of tunneling through an outermost 
BiO layer in BSSCO compounds which has important consequences for the
form factor of substitutional impurities in the CuO$_{2}$-plane as was
pointed out by Martin, Balatsky and Zaanen \cite{martin}.   
Returning to the case of
Ca$_{2-x}$Na$_{x}$CuO$_{2}$Cl$_{4}$ the injection of an
electron (or hole) through the 3$p_{z}$ orbital of a Cl ion
neighboring a   Na-ion leads to finite amplitudes for injection into
the 4 Cu sites which are nearest neighbors (nn) to the Cu underneath
the Cl ion.  Of these 4 sites 2 are nn sites of the Na  ion and 2 are
nnn sites (see Fig.~1).

The next question that arises concerns the tunneling probability
is whether the tunneling amplitude into these 4 sites should be
added coherently or incoherently.  Consider the case of positive
bias or electron injection from the tip.  Clearly the Coulomb
attraction between a hole and the Na-acceptor 
leads to increased occupation of states centered on the 4 n.n.
Cu sites and to a lesser extent on the 8 n.n.n. sites.  
Explicit form of hole wavefunction around a dopant was 
numerically calculated for the case of Li$^{+}$ dopant 
in ladder system \cite{lauchli}. 
Similar calculation is possible for this case 
but this is beyond the scope of this paper. 
This in turn enhance the hole density on these sites.  
When we 
calculate the electron spectral function $A(\mathbf{x},\omega)$ that
determines the conductance 
\begin{eqnarray}
A(\mathbf{x},\omega) &=& \sum_{\alpha, \sigma} 
\Bigl| \Bigl\langle \Psi_{\alpha}^{\rm 0h} \Big| 
\sum_{\mathbf{x}'} h (\mathbf{x},\mathbf{x}') c_{\mathbf{x}',\sigma}^\dagger 
\Big| \Psi_{0}^{\rm 1h} \Bigr\rangle \Bigr|^2
\nonumber\\
&&\ \times \delta\left( \omega - E_{\alpha}^{\rm 0h} 
                       + E_{0}^{\rm 1h} \right) . 
\label{eq:spectral}
\end{eqnarray}
We are interested in an energy range stretching up to several hundred
meV.  Therefore the relevant matrix element will be a incoherent 
superpositions of tunneling processes into the 4 Cu sites which are
n.n. to the Cl site.   Here $|\Psi_{0}^{\rm 1h} \rangle$ is the one-hole
groundstate with energy $E_{0}^{\rm 1h}$ and 
$|\Psi_{\alpha}^{\rm 0h} \rangle$ 
are eigenstates of a stoichiometric system, and $h(\mathbf{x},\mathbf{x}')$ 
is the amplitude of the hole wave function centered on the Cu-site 
$\mathbf{x}'$, at the Cl site at position, $\mathbf{x}$.  
For calculating the spectral function, we average the energy 
over the allowed range for the injected electron so that 
we may treat the sum over the one-hole eigenstates as a sum 
over a complete set of states.  
Under these incoherent conditions the 
$ \left\langle A(\mathbf{x},\omega) \right\rangle_{\omega\mathrm{-av}} 
\sim \sum_{\mathbf{x}'} 
h^2 (\mathbf{x},\mathbf{x}') n(\mathbf{x}')$ 
where $n(\mathbf{x}')$ is the hole density at $\mathbf{x}'$. 
In the neighborhood of a Na$^{+}$-ion the Coulomb attraction leads to an 
enhanced hole density on the inner ring of 4 nearest neighbor sites and 
to a lesser extent on the outer ring of 8 next nearest neighbor sites 
(see Fig.~1).  This in turn will lead to an enhanced tunneling signal on 
the inner ring ($f_1$) and the outer ring ($f_2$).  The Fourier transformed 
cluster form factor is then 
\begin{eqnarray}
  f(\mathbf{k})&=&4f_1 \cos \frac{k_x}{2} \cos \frac{k_y}{2}
  \nonumber\\
  &&+ 4 f_2 \left[ \cos \frac{3k_x}{2} \cos \frac{ k_y}{2}
                +\cos \frac{ k_x}{2} \cos \frac{3k_y}{2} \right] . 
\label{eq:fk}
\end{eqnarray}
We do not attempt detailed calculations
of $f_1$, and $f_2$ but treat them as phenomenological parameters.  Note
in this letter we will concentrate on electron injection only and so
our calculations are for $A(\bar{\omega})$, 
where $\bar{\omega}$ is an energy of order 100 [meV] or above, 
where the experimental data do not depend on the energy so much.  
Strictly speaking this should be determined from experiment by
normalizing the spectra in the -ve bias (hole injection) side rather
than the +ve bias side chosen by Hanaguri et al.\cite{martin}. 
An explicit proposal along these lines has been 
made by Randeria and coworkers \cite{randeria}. 

\begin{figure}[b]
\begin{center}
\end{center}
\caption{Typical configuration of Na ion distribution
simulated by Monte Carlo calculations.  The system size
is 100$\times$100 sites and the Na ion
density is 0.10.    
Strength of elastic component of Na-Na interaction
is (a) A=0.5, and (b) A=1.0.}
\label{fig:conf} 
\end{figure}

The STM signal intensity is given by
\begin{equation}
I(\mathbf{x}) = \int d^2 \mathbf{R} \, 
\rho(\mathbf{R}) f(\mathbf{x}-\mathbf{R}) 
\label{eq:intensity}
\end{equation}
where $f(\mathbf{x}-\mathbf{R})$ is the intensity
of a cluster at Cl site  $\mathbf{x}$ around a Na-acceptor at $\mathbf{R}$ 
and it is the inverse Fourier transform of $f(\mathbf{k})$ in 
Eq.~(\ref{eq:fk}).  
$\rho(\mathbf{R})$ is the distribution function for the Na-ions.  
At the relevant densities $x \approx 0.1$ the
concentration of the Na ions in the outermost 
Ca$_{2-x}$Na$_{x}$CuO$_{2}$Cl$_{2}$ 
layer is not in the dilute regime so that it is possible that the
Na-Na interactions lead to short range order in the positions of  the
Na acceptors.  To estimate the effects of these interactions on the
distribution we simulated the annealing of an initial random
2-dimensional distributions through inter-ion interactions using a
Monte Carlo algorithm.  The interaction has a Coulomb repulsion
component and an elastic  component 
\begin{equation}
  \label{eq:elastic}
  V (\mathbf{R}) = \frac{1}{R} - A \frac{\cos 4\theta}{R^2}.  
\end{equation}
The former is radially symmetric but the second term is in general
anisotropic.  We estimate it following Eshelby\cite{esherby} 
for a 2-dimensional
square lattice and we ignore effects associated with
the fact that the Na-acceptors lie both above and below the topmost
CuO-plane.  
Here $\theta$ is the angle of vector $\mathbf{R}$ measured 
from the principal axis of the lattice. The intensity 
$A=2\pi c^2 \eta (\lambda+\mu)/(\lambda+2\mu)$ is 
given by Lam\'e constants, $\lambda =\lambda_{xx,yy}$ and 
$\mu = \lambda_{xy,xy}$, and 
the tetragonal anisotropy, $\eta = \lambda_{xx,xx}-(\lambda+2\mu)$, 
with $c$ being a constant proportional to local volume 
change due to impurity doping.  
This is normalized such that the coefficient of the Coulomb 
term is set to be unity.  

\begin{figure}[tb]
\begin{center}
\end{center}
\caption{(a) Fourier transform of Na-ion distribution,  
$\langle |\rho (\mathbf{k})|^2 \rangle$, 
annealed for the potential $V(\mathbf{R})$ with $A$=1.0.  
Ensemble average is taken over 
$10^2$ configurations generated by Monte Carlo calculations 
and the results are smoothed by convoluting with 
Gaussian function with width 0.03$\pi$.
The system size is 100$\times$100 sites and the Na ion
density is 0.10.
(b) Fourier transform of random initial configurations 
at the same density before annealing.}
\label{fig:Fourier} 
\end{figure}

We have investigated stable configuration of doped Na$^+$-ions 
with the above interaction potential.  For this simulated annealing, 
we have used a Monte Carlo algorithm.  Starting from a random 
initial configuration, we choose a Na-ion in turn and find 
a vacant site randomly.  We then calculate how much 
the interaction energy changes, $\Delta E$, if the chosen 
ion is to move.  If $\Delta E <0$, we always move 
that ion to achieve a stabler configuration, 
whereas even if $\Delta E >0$, we accept the new configuration 
with the Boltzmann probability $e^{-\Delta E/T}$ at temperature $T$. 
This update process is applied to all the ions in turn, and 
the procedure is repeated many times to obtain a stable ion 
configuration.  
Typically, we have set the 
temperature $T= 0.01 - 0.05$, and repeated the Monte Carlo procedure 
$10^3 - 10^4$ steps for each initial configuration.  
We have checked that obtained results do not depend on these 
parameters sensitively.   
In Fig.~2  we show  typical distributions resulting
from Monte Carlo annealing of samples with $10^4$ sites and $10^3$ 
ions.  Periodic boundary conditions are imposed to 
both directions and the interaction $V(\mathbf{R})$ is tailored 
for the boundary conditions by using the mirror charge technique.  

\begin{figure}[tb]
\begin{center}
\end{center}
\caption{Structure factor of holes localizing around 
doped Na ions, $\langle |I(\mathbf{k})|^2 \rangle$.
Ensemble average is taken over 
$10^2$ configurations generated by Monte Carlo calculations 
and the results are smoothed by convoluting with 
Gaussian function with width 0.03$\pi$.  
The system size is 100$\times$100 sites and the Na ion
density is 0.10.    
Coupling constant is (a) A=0.5, (b) A=1.0, and (c) A=0.0.}
\label{fig:Sk} 
\end{figure}

Once the annealed configuration is obtained,  
we calculate the structure factor of Na ions by taking 
the Fourier transform. 
Result for 10\% doping and $A$=1.0 is plotted in Fig.~3(a), and 
the Fourier transform of random initial configurations 
is also shown in Fig.~3(b) in comparison. 
We have also checked that ensemble average over a larger 
set of samples (e.g., $10^3$) gives essentially the same results.   
The Na-ion structure factor is 
averaged over initial configurations, and then smoothed 
by Gaussian broadening with width $0.03\pi$.  The final structure 
factor is calculated by multiplying it with the form factor 
of hole density localizing around Na ions 
\begin{equation}
  \left| I(\mathbf{k})\right|^2 = 
  \left| \rho (\mathbf{k})\right|^2  \left| f(\mathbf{k})\right|^2 .  
\end{equation}
In Fig.~4 (a) and (b), we show representative sets of data for two 
values of the
relative strength of the elastic and Coulomb interactions.  
The data for the case of pure Coulomb interaction 
are also shown in Fig.~4(c) in comparison.  
We see that the main characteristic of the checkerboard pattern observed in
the experiments,\cite{martin} namely peaks at
wavevectors $(\pi/2,0)$ and $(0,\pi/2)$ are nicely reproduced 
when the anisotropic elastic energy is considered $A>0$,
especially for the case with stronger elastic interactions.
The stronger tendency
for the Na-ions to line up along the crystal axes in these
simulations is reflected in the structure factor by the suppression
of weight along the diagonal leading to the peaks
being concentrated near the vertical and horizontal axes.

Our calculations show that at least the broad features of the data
can be reproduced by our two assumptions,
enhanced hole density around the acceptor and 
short range correlations in the positions of the
Na-acceptors.  Such short range order necessarily induces a
corresponding modulation in the hole distribution. In addition the
hole density will be  distributed over a cluster of sites around
each acceptor.  Both effects influence the final structure factor.
The  STM data contains a lot of more detailed spectroscopic
information, e.g. small peaks in the conductance.  These features are
beyond the scope of the calculations presented here.  The STM
experiments give us really detailed information on an unprecedented
scale but for now at least, we can only  analyze the main features.
A key point in our model is the enhanced hole density in the
vicinity of an acceptor which hopefully can be explicitly tested 
in experiment.

\begin{acknowledgement}
  This work started when TMR was staying at the Yukawa Institute.
TMR is grateful for the visiting position at the Yukawa Institute.
The authors thank T.~Hanaguri and S.~Davis for discussions 
of their experiments.  
HT is supported by a Grant-in-Aid from the Ministry of Education,
Science, Sports, and Culture of Japan.  
\end{acknowledgement}

%
%

\end{document}